\documentclass[aps,nofootinbib,prd,preprint,preprintnumbers]{revtex4}
\usepackage{graphicx}

\usepackage{amsmath}
\usepackage{amssymb}

\usepackage{hyperref}
\usepackage{color}

\def\vev#1{\left\langle #1\right\rangle}
\def\SM{$\mathrm{SU(3)_c \otimes SU(2)_L \otimes U(1)_Y}$ }
\def\21{$\mathrm{SU(2)_L \otimes U(1)_Y}$ }
\def\lfv{lepton flavour violation }
\def\lnv{lepton number violation }

\newcommand{\sm}{standard model }

\newcommand{\AddrAHEP}{AHEP Group, Institut de F\'{i}sica Corpuscular --
  C.S.I.C./Universitat de Val\`{e}ncia, Parc Cientific de Paterna.\\
  C/Catedratico Jos\'e Beltr\'an, 2 E-46980 Paterna (Val\`{e}ncia) - SPAIN}

\newcommand{\Virginia}{Center for Neutrino Physics, Virginia Tech,
  Blacksburg, VA 24061, USA}

\newcommand{\Cinvestav}{Departamento de F\'{\i}sica, Centro de
  Investigaci{\'o}n y de Estudios Avanzados del IPN\\ Apdo. Postal
  14-740 07000 Mexico, DF, Mexico}
  
\begin{document}

\preprint{IFIC/15-14}

\title{On the description of non-unitary neutrino mixing}

\author{F. J. Escrihuela~$^1$}\email{franesfe@alumni.uv.es}
\author{D. V. Forero~$^2$}\email{dvanegas@vt.edu}
\author{O. G. Miranda~$^3$}\email{omr@fis.cinvestav.mx}
\author{M. T\'ortola~$^1$}\email{mariam@ific.uv.es}
\author{J. W. F. Valle~$^1$} \email{valle@ific.uv.es, URL:
  http://astroparticles.es/} 
\affiliation{$^1$~\AddrAHEP}
\affiliation{$^2$~\Virginia}
\affiliation{$^3$~\Cinvestav}

\begin{abstract}
  Neutrino oscillations are well established and the relevant
  parameters determined with good precision, except for the CP phase,
  in terms of a unitary lepton mixing matrix. Seesaw extensions of the
  Standard Model predict unitarity deviations due to the admixture of
  heavy isosinglet neutrinos.
  We provide a complete description of the unitarity and universality
  deviations in the light neutrino sector.
  Neutrino oscillation experiments involving electron or muon
  neutrinos and anti-neutrinos are fully described in terms of just
  three new real parameters and a new CP phase, in addition to the
  ones describing oscillations with unitary mixing.
  Using this formalism we describe the implications of non-unitarity
  for neutrino oscillations and summarize the model-independent
  constraints on heavy neutrino couplings that arise from current
  experiments.
\end{abstract}


\maketitle
\section{Introduction}

\begin{figure}[htb] \centering
   \includegraphics[width=.4\linewidth]{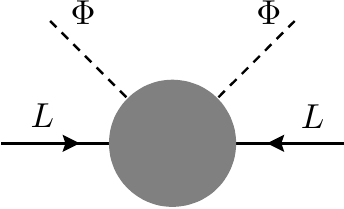}
    \caption{\label{fig:d5} 
    Dimension five  operator responsible for neutrino mass.}
\end{figure}

Neutrino masses, without which current neutrino oscillation data can
not be understood~\cite{Tortola:2012te}, are here to
stay~\cite{Nunokawa:2007qh}.  It has been long noted that small
neutrino masses can arise from an effective \lnv dimension-five
operator $\mathcal{O}_5 \propto L L \Phi \Phi~,$ which may arise from
unknown physics beyond that of the \SM model. Here $L$ denotes one of
the three lepton doublets and $\Phi$ is the \sm scalar
doublet~\cite{Weinberg:1980bf}.
After electroweak symmetry breaking takes place through the nonzero
vacuum expectation value (vev) $\vev{\Phi}$ such operator leads to
Majorana neutrino masses. In contrast to the charged fermion masses,
which arise directly from the coupling of the scalar Higgs, neutrino
masses appear in second order in $\vev{\Phi}$ and imply \lnv by two
units ($\Delta L=2$) at some large scale.
This fact accounts for the smallness of neutrino masses relative to
those of the \sm charged fermions.
This is all we can say from first principles about the operator
$\mathcal{O}_5$ in Fig.~\ref{fig:d5}. In general we have no clue on
the {\sl mechanism} giving rise to $\mathcal{O}_5$, nor its associated
mass {\sl scale}, nor the possible details of its {\sl flavour
  structure}.

\begin{table*}[!h]
  \centering
\begin{math}
\begin{array}{|c|c|} \hline
& \ \ \ {\mbox SU(3)\otimes SU(2)\otimes U(1)} \\
\hline
L_a = (\nu_a, l_a)^T & (1,2,-1/2)\\
e_a^c   & (1,1,1)\\
\hline
Q_a = (u_a, d_a)^T    & (3,2,1/6)\\
u_a^c   & (\bar{3},1,-2/3)\\
d_a^c   & (\bar{3},1,1/3)\\
\hline
\Phi  & (1,2,1/2)\\
\hline
\end{array}
\end{math}
\caption{Matter and scalar multiplets of the Standard Model.}
\label{tab:SM}
\end{table*}

One may assume that $\mathcal{O}_5$ is induced at the tree level by
the exchange of heavy ``messenger'' particles, whose mass lies at a
scale associated to the violation of the global lepton number symmetry
by new physics, beyond that of the \SM model,
$$ m_{\nu} = \lambda_0 \frac{\vev{\Phi}^2}{M_X}, $$
where $\lambda_0$ is some unknown dimensionless constant.
For example gravity, which in a sense "belongs" to the SM, could
induce the dimension-five \lnv operator
$\mathcal{O}_5$~\cite{Giddings:1988cx,Banks:2010zn}.  In such a
minimalistic scenario~\cite{deGouvea:2000jp} the large scale $M_X$ in
the denominator is the Planck scale and hence the neutrino mass that
results is too small to account for current neutrino oscillation data.
Hence we need genuine ``new physics'' in order to generate neutrino
masses this way. 

Neutral heavy leptons (NHL) arise naturally in several extensions of
the Standard Model. Their possible role as messengers of neutrino mass
generation constitutes one of their strongest motivations and a key
ingredient of the type-I seesaw
mechanism~\cite{Valle:2015pba,Schechter:1980gr,Schechter:1981cv,gell-mann:1980vs,mohapatra:1981yp}
in any of its variants. If realized at the Fermi
scale~\cite{Mohapatra:1986bd,GonzalezGarcia:1988rw,Akhmedov:1995ip,Akhmedov:1995vm,Malinsky:2005bi,Boucenna:2014zba,Dev:2009aw,Dev:2013oxa,Drewes:2015iva},
it is likely that the ``seesaw messengers'' responsible for inducing
neutrino masses would lead to a variety of phenomenological
implications. These depend on the assumed gauge structure. Here for
definiteness and simplicity, we take the minimal \SM structure which
is well tested experimentally. In this case one can have, for example,
\begin{enumerate}
\item Light isosinglet leptons are usually called ``sterile''. If they
  lie in the eV range they could help accommodate current neutrino
  oscillation anomalies~\cite{Aguilar:2001ty,Aguilar-Arevalo:2013pmq}
  by taking part in the oscillations.  Sterile neutrinos at or above
  the keV range might show as distortions in weak decay
  spectra~\cite{Shrock:1980ct} and be relevant for
  cosmology~~\cite{Asaka:2005an}.
\item Heavy isosinglet leptons below the Z mass could have been seen
  at LEP I~\cite{Dittmar:1989yg,Akrawy:1990zq,Abreu:1996pa}. Likewise,
  TeV NHLs might be seen in the current LHC experiment, though in the
  latter case rates are not expected to be large in the \SM theory.
\item Whenever NHL are too heavy to be emitted in weak decay.
  processes, the corresponding decay rates would decrease, leading to
  universality violation~\cite{Gronau:1984ct}.
\item The admixture of NHL in the charged current weak interaction
  would affect neutrino oscillations, since they would not take part
  in oscillations.  These would be effectively described by a
  non-unitary mixing matrix~\cite{Forero:2011pc}.
\item If Majorana-type, NHL would modify rates for \lnv processes such
  as neutrinoless double beta ($0\nu\beta\beta$) decays through
  long-range (mass mechanism), as well as induce short-range
  contributions~\cite{Schechter:1981bd,Hirsch:2006tt,Rodejohann:2011mu}.
\item NHL would induce charged \lfv
  processes~\cite{Bernabeu:1987gr,Forero:2011pc}. However the
  corresponding restrictions depend on very model-dependent rates.
\end{enumerate}
   
In what follows we consider the generic structure of the lepton mixing
matrix relaxing the unitarity approximation~\footnote{In sections
  II-VI we mainly consider isosinglet neutrinos above 100 GeV or so,
  hence too heavy to take part in oscillations or low energy weak
  decay processes.}. We show that their most general form is
factorizable, so that current experiments involving only electron and
muon neutrinos or anti-neutrinos can be effectively described in terms
of just three new real parameters and one new CP violation phase.
We illustrate how these parameters affect oscillations and discuss the
main restrictions on such generalized mixing structure that follow
from universality tests. For logical completeness we also present a
brief compilation of various model--independent constraints on NHL
mixing parameters within the same parametrization, including those
that follow from the possibility of direct NHL production at high
energy accelerator experiments.

\section{The formalism}

Isosinglet neutral heavy leptons couple in the weak charged current
through mixing with the standard isodoublet neutrinos. The most
general structure of this mixing matrix has been given in the
symmetric parametrization in Ref.~\cite{Schechter:1980gr}. Here we
consider an equivalent presentation of the lepton mixing matrix which
manifestly factorizes the parameters associated to the heavy leptons
from those describing oscillations of the light neutrinos within the
unitarity approximation. Here we present its main features, details
are given in the appendix~\footnote{We consider stable neutrinos,
  neutrino decays were discussed, for instance, in
  Ref.~\cite{Berryman:2014yoa}.}.

For the case of three light neutrinos and $n-3$ neutral heavy leptons,
one can break up the matrix $U^{n\times n}$ describing the
diagonalization of the neutral mass matrix
as~\cite{Hettmansperger:2011bt}
\begin{equation}
U^{n\times n}=\left(\begin{array}{cc} N & S\\
V & T
\end{array}\right)\label{eq:ULindner_C1} ,
\end{equation}
where $N$ is a $3\times3$ matrix in the light neutrino sector, while
$S$ describes the coupling parameters of the extra isosinglet states,
expected to be heavy (for a perturbative expansion for $U^{n\times n}$
see~\cite{Schechter:1981cv}).  As shown in the appendix, the matrix
$N$ can be expressed most conveniently~\footnote{There are other forms
  for the light-neutrino mixing matrix, where the pre-factor
  off-diagonal zeroes are located at different entries. However
  Eq.~(\ref{eq:Ndescopm_C1}) is the most convenient to describe
  current neutrino experiments.} as
\begin{equation}
N=N^{NP}\, U=\left(\begin{array}{ccc}\alpha_{11} & 0 & 0\\
\alpha_{21} & \alpha_{22} & 0\\
\alpha_{31} & \alpha_{32} & \alpha_{33}
\end{array}\right)\: U
\label{eq:Ndescopm_C1} ,
\end{equation}
where $U$ is the usual unitary form of the $3\times 3$ leptonic mixing
matrix probed in neutrino oscillation studies~\footnote{As discussed
  in Ref.~\cite{Rodejohann:2011vc}, this may, for example, be
  parameterized in the original symmetric way or equivalently as
  prescribed in the Particle Data Group.} corrected by the left
triangle pre-factor matrix, $N^{NP}$, characterizing unitarity
violation.

Note that Eq.~(\ref{eq:Ndescopm_C1}) provides a most convenient,
general and complete description of the propagation of solar,
atmospheric and terrestrial neutrinos from reactors, radioactive
sources and accelerators beams, relaxing the unitarity
approximation. Due to the zeroes in the first two rows of the
pre-factor matrix in Eq.~(\ref{eq:Ndescopm_C1}) it is clear that the
only extra parameters beyond those characterizing unitary mixing are
four: the two real parameters $\alpha_{11}$ and $\alpha_{22}$ plus the
complex parameter $\alpha_{21}$ which contains a single CP
phase. Indeed the existence~\cite{Branco:1989bn} and possible
effects~\cite{Rius:1989gk} extra CP phases associated to the admixture
of NHL in the charged leptonic weak interaction had already been noted
in the early paper in~\cite{Schechter:1980gr}. The new point here is
that, despite the proliferation of phase parameters, only one
combination enters the ``relevant'' neutrino oscillation
experiments. This holds irrespective of the number of extra heavy
isosinglet neutrino states present. Other studies, such
as~\cite{Antusch:2006vwa,Xing:2007zj,Xing:2011ur}, appear as
particular cases with a fixed number of extra heavy isosinglet
neutrino states, any of which can be expressed in terms of the same
set of parameters $\alpha_{ij}$.  Similarly, the matrix U may be
expressed in different ways, such as in PDG form or in our fully
symmetric description, particularly useful for phenomenological
analyses.
The diagonal elements, $\alpha_{ii}$, are real and expressed in a
simple way as
\begin{eqnarray}
\alpha_{11} \: &=& \: c_{1\, n}\: c_{\,1n-1}c_{1\, n-2}\ldots c_{14}  , \nonumber \\
\alpha_{22} \: &=& \: c_{2\, n}\: c_{\,2n-1}c_{2\, n-2}\ldots c_{24} , \\
\alpha_{33} \: &=& \: c_{3\, n}\: c_{\,3n-1}c_{3\, n-2}\ldots c_{34}  \nonumber ,
\end{eqnarray}
in terms of the cosines of the mixing
parameters~\cite{Schechter:1980gr}, ${c}_{ij}=\cos\theta_{ij}$.

Now the off-diagonal terms $\alpha_{21}$ and $\alpha_{32}$ are
expressed as a sum of $n-3$ terms
\begin{eqnarray}
\alpha_{21} \: = \:
    c_{2\, n}\: c_{\,2n-1}\ldots c_{2\, 5}\:{\eta}_{24}\bar{\eta}_{14}\: +\: 
    c_{2\, n}\: \ldots c_{2\, 6}\:{\eta}_{25}\bar{\eta}_{15}\:c_{14} +\: 
    \ldots\:+ {\eta}_{2n}\bar{\eta}_{1n}\:c_{1n-1}\:c_{1n-2}\:\ldots\:c_{14} 
\nonumber \, , \\
\alpha_{32} \: = \:
    c_{3\, n}\: c_{\,3n-1}\ldots c_{3\, 5}\:{\eta}_{34}\bar{\eta}_{24}\: +\: 
    c_{3\, n}\: \ldots c_{3\, 6}\:{\eta}_{35}\bar{\eta}_{25}\:c_{24} +\: 
    \ldots\:+ {\eta}_{3n}\bar{\eta}_{2n}\:c_{2n-1}\:c_{2n-2}\:\ldots\:c_{24} ,  
\label{eq:alfa_crossed_C1}
\end{eqnarray}
where ${\eta}_{ij}=e^{-i\phi_{ij}}\,\sin\theta_{ij}$ and its
conjugate $\bar{\eta}_{ij}=-e^{i\phi_{ij}}\,\sin\theta_{ij}$ contain all
of the CP violating phases.  Finally, by neglecting quartic terms in
$\sin\theta_{ij}$, with $j=4,5,\cdots$ one finds a similar expression
for $\alpha_{31}$,
\begin{eqnarray}
\alpha_{31} \: = \:
    c_{3\, n}\: c_{\,3n-1}\ldots c_{3\, 5}\:{\eta}_{34} c_{2\, 4} \bar{\eta}_{14}\: +\: 
    c_{3\, n}\: \ldots c_{3\, 6}\:{\eta}_{35} c_{2\, 5}\bar{\eta}_{15}\:c_{14} +\: 
    \ldots\: \nonumber \\
    + \: {\eta}_{3n} c_{2\,
  n}\bar{\eta}_{1n}\:c_{1n-1}\:c_{1n-2}\:\ldots\:c_{14}  \, .
\label{eq:alfa_31}
\end{eqnarray}
In summary, by choosing a convenient ordering for the products of the
complex rotation matrices $\omega_{ij}$ (see appendix), one obtains
a parametrization that separates all the information relative to the
additional leptons in a simple and compact form, containing three
zeroes. We will now concentrate on this specific parametrization.

\section{Non-unitary neutrino mixing  matrix}

Given the above considerations and the chiral nature of the \SM model,
we notice that the couplings of the $n$ neutrino states in the charged
current weak interaction can be described by a rectangular
matrix~\cite{Schechter:1980gr}
\begin{equation}
K=\left(\begin{array}{cc} N & S
\end{array}\right)\label{eq:K} ,
\end{equation}
with $N$ a $3\times 3$ matrix described by Eq.~(\ref{eq:Ndescopm_C1})
and $S$ a $3\times (n-3)$ matrix. This can be parametrized in the
symmetric form or as prescribed in the Particle Data Group. The
relative pros and cons of the two presentations are considered in
Ref.~\cite{Rodejohann:2011vc}.

The presence of extra heavy fermions that mix with the active light
neutrinos would imply the effective non-unitarity of the $3\times3$
light neutrino mixing matrix, hence modifying several SM observables.
For example, note that the unitarity condition will take the form
\begin{equation}
KK^\dagger = NN^\dagger + SS^\dagger = I ,
\label{eq:unita}
\end{equation}
with 
\begin{equation}
NN^\dagger=\left(\begin{array}{lcccl} \alpha_{11}^2\ & & 
                                 \alpha_{11}\alpha^*_{21} &  &
                                 \alpha_{11}\alpha^*_{31} \\
      \alpha_{11}\alpha_{21} & &
                    \alpha_{22}^2  + |\alpha_{21}|^2 & &
                    \alpha_{22}\alpha^*_{32} + \alpha_{21}\alpha^*_{31}  \\
      \alpha_{11}\alpha_{31} & &
                    \alpha_{22}\alpha_{32} + \alpha_{31}\alpha^*_{21} & &
     \alpha_{33}^2 + |\alpha_{31}|^2 + |\alpha_{32}|^2 
\end{array}\right) \, . \label{eq:NN}
\end{equation}

We will show that, with the parametrization discussed here, one can,
at least in principle, introduce all of the information of the extra
$n-3$ states into the $\alpha_{ij}$ parameters in a simple compact
form. The method is completely general and includes all the relevant
CP phases.
In what follows we will consider different direct or indirect tests of
the existence of the extra heavy fermions, expressing the relevant
observables in terms of these parameters, in order derive the relevant
constraints.

\section{Universality constraints}

First one notes that if, as generally expected due to their gauge
singlet nature, the heavy leptons can not be kinematically emitted in
various weak processes such as muon or beta decays, these decays will
be characterized by different effective Fermi constants, hence
breaking universality.
One can now apply the above formalism in order to describe the various
weak processes and to derive the corresponding experimental
sensitivities. We first discuss the universality constraint, already
reported in the
literature~\cite{Gronau:1984ct,Nardi:1994iv,Langacker:1988ur,GonzalezGarcia:1990fb,Abada:2012mc,Abada:2013aba,Atre:2009rg,Abada:2014nwa,Antusch:2014woa},
in order to cast it within the above formalism.
Comparing muon and beta decays one finds
\begin{equation}\label{eq1}
G_{\mu}=G_F\, \sqrt{(NN^\dagger)_{11}(NN^\dagger)_{22}}\, 
= G_F\, \sqrt{\alpha_{11}^2(\alpha_{22}^2+|\alpha_{21}|^2)},
\end{equation}
and
\begin{equation}\label{eq1A}
G_{\beta}=G_F\, \sqrt{(NN^\dagger)_{11}}\, 
= G_F\, \sqrt{\alpha_{11}^2}.
\end{equation}
Therefore, all the observables related to Fermi constant will be
affected by this change, for instance, the quark CKM matrix
elements~\cite{Nardi:1994iv}.  In particular, the CKM matrix elements
$V_{ud}$ and $V_{us}$ are proportional to $G_\mu$. These matrix
elements are measured in $\beta$-decay, $K_{e3}$ decay, and hyperon
decays. The effect on $G_\mu$, therefore, modifies $V_{ui}$ and the
unitarity constraint for the first row of the CKM is now expressed
as~\cite{Nardi:1994iv,Langacker:1988ur}:
\begin{equation}\label{eq3}
\sum_{i=1}^{3} |V_{ui}|^2=
\left(\frac{G_\beta}{G_\mu}\right)^2 =
\left(\frac{G_F\sqrt{(NN^\dagger)_{11}}}{G_F\sqrt{(NN^\dagger)_{11}(NN^\dagger)_{22}}}\right)^2
=\frac{1}{(NN^\dagger)_{22}},
\end{equation}
where the Eq.~(\ref{eq1})  has been used in the last equality.
Following the previous equation one gets~\cite{Agashe:2014kda}: 
\begin{equation}\label{eq11}
\sum_{i=1}^{3} |V_{ui}|^2=\frac{1}{\alpha_{22}^2 + |\alpha_{21}|^2}
=0.9999\pm 0.0006,
\end{equation}
and, therefore, $1-(NN^\dagger)_{22} = (SS^\dagger)_{22} = 1 -
  \alpha_{22}^2-|\alpha_{21}|^2 <0.0005$
  at 1$\sigma$. 

  There are other universality tests that give constraints on these
  $\alpha$ parameters.  For example, universality implies that the
  couplings of the leptons to the gauge bosons are flavor independent,
  a feature that emerges in the the \sm without heavy leptons.
  In the presence of heavy isosinglets, these couplings will be flavor
  dependent; the ratios of these couplings can be extracted from weak
  decays and they are expressed as~\cite{Nardi:1994iv}:
\begin{equation}\label{eq2}
\left(\frac{g_a}{g_\mu} \right)^2=
\frac{(NN^\dagger)_{aa}}{(NN^\dagger)_{22}} 
\quad a=1,3 \, .
\end{equation}

\begin{figure}
\centerline{
\includegraphics[scale=0.7]{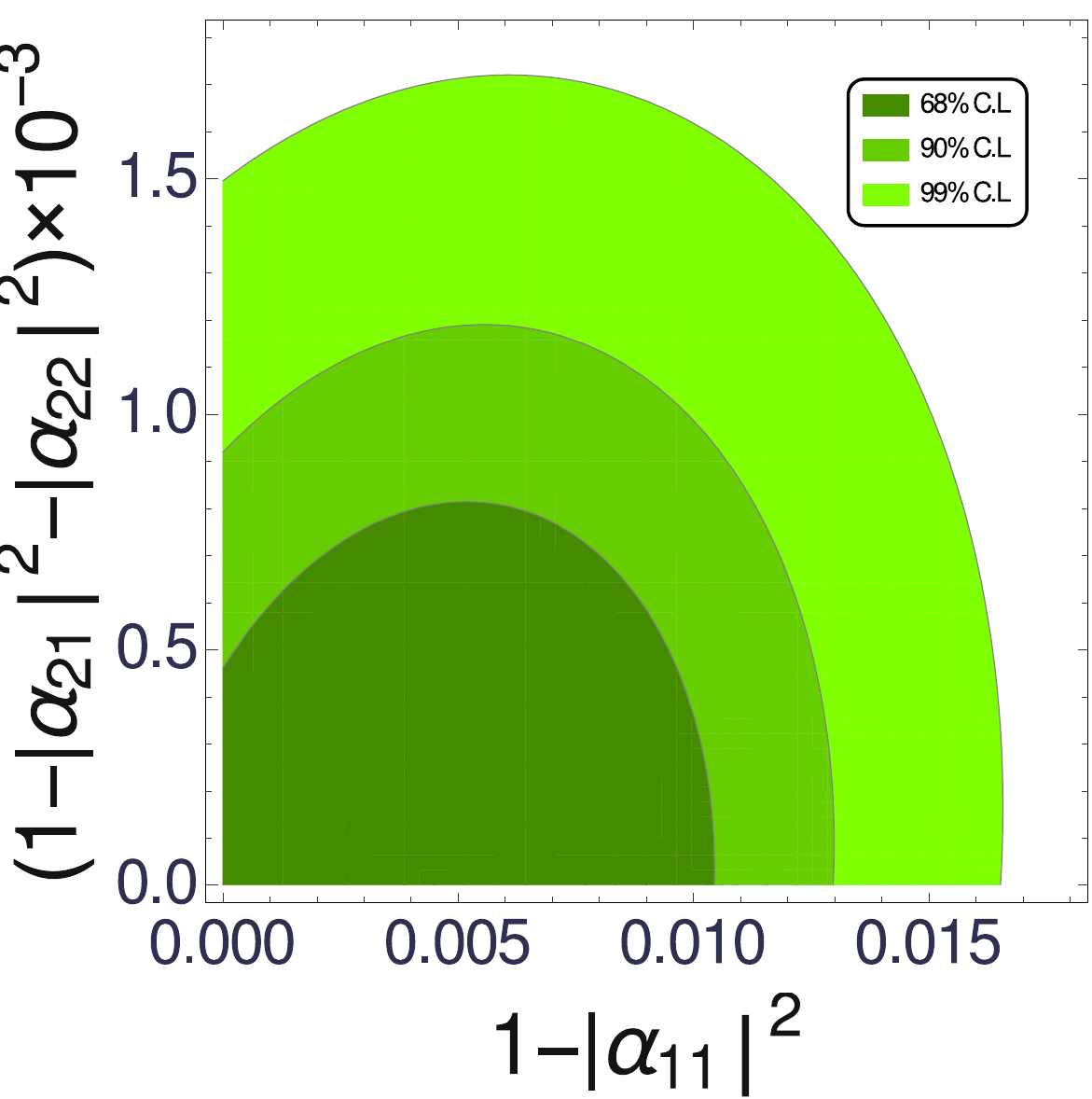}
}
\caption{\label{fig:universality} Constraints on the deviations from
  unitarity.}
\end{figure}
For $a=1$, this ratio can be constrained by comparing the experimental
measurement and the theoretical prediction of the pion decay branching
ratio~\cite{Abada:2012mc}:
\begin{equation}
R_{\pi} = \frac{\Gamma (\pi^+ \to e^+\nu)}{\Gamma (\pi^+ \to \mu^+\nu)}.
\end{equation}
One obtains~\cite{Abada:2012mc,Czapek:1993kc}:
\begin{equation}\label{eq10}
r_\pi= \frac{R_{\pi}}{R^{SM}_{\pi}} =
\frac{(NN^\dagger)_{11}}{(NN^\dagger)_{22}} =
\frac{\alpha_{11}^2}{\alpha_{22}^2 + |\alpha_{21}|^2} =
\frac{(1.230\pm 0.004)\times 10^{-4}}{(1.2354\pm 0.0002)\times 10^{-4}}=0.9956\pm0.0040
\end{equation} 
which implies $1-\alpha_{11}^2<0.0084$ at 1$\sigma$ for the least
conservative case of $\alpha_{22}^2 + |\alpha_{21}|^2=1$.  This
procedure was adopted in Ref.~\cite{Atre:2009rg}.
However, in general $[(NN^\dagger)_{22}] \ne 1$, and it can be
estimated using the unitarity constraints on the CKM matrix discussed
above. Combining both constraints (from Eqs.~(\ref{eq11}) and
(\ref{eq10})) we obtain the results shown in Figure
\ref{fig:universality}, restricting the parameter combinations shown
in the plot. These translate in the constraints
\begin{eqnarray}
1-\alpha_{11}^2 &<& 0.0130 \nonumber \, , \\
1-\alpha_{22}^2-|\alpha_{21}|^2 &<& 0.0012 \, ,
\label{eq:universality-bounds}
\end{eqnarray}
at $90\%$ C.L. for 2 d.o.f. One can make use of a third observable
in order to have constraints for every independent parameter. This
will be discussed in the next section.

For the sake of completeness we now show the constraints coming from
the $\mu-\tau$ universality which, using Eq.~(\ref{eq2}), give the
bound:
\begin{equation}\label{eq12}
\frac{(NN^\dagger)_{33}}{(NN^\dagger)_{22}}=0.9850 \pm 0.0057 \,.
\end{equation}
This implies $1-(NN^\dagger)_{33}=(SS^\dagger)_{33}<0.0207$ at
1$\sigma$ for the least conservative case of
$(SS^\dagger)_{22}=0$. The experimental value was taken from
Ref.~\cite{Aubert:2009qj}. We now turn to neutrino oscillations.

\section{non-unitarity effect on neutrino oscillations}

In this section we focus on neutrino oscillation experiments.  First
we obtain general expressions for neutrino survival and conversion
probabilities in this parametrization and confront them with the
existing experimental data. The general expressions will be relatively
simple, especially if we neglect cubic products of $\alpha_{21}$,
$\sin\theta_{13}$, and $\sin(\frac{\Delta m^2_{21}}{4E})$, which is a
reasonable approximation for many applications. The results of this
approach for the three probabilities discussed in this section are
shown in Eqs. (\ref{eq:Pmue}),  (\ref{eq:Pmumu3}) and (\ref{eq:Pee}).

For the case of the muon neutrino conversion probability into electron
neutrino we have:
\begin{eqnarray}
P_{\mu e} = \sum^3_{i,j} N^*_{\mu i}N_{ei}N_{\mu j}N^*_{ej} &-& 
        4 \sum^3_{j>i} Re\left[
        N^*_{\mu j}N_{ej}N_{\mu i}N^*_{ei}\right] 
        \sin^2\left(\frac{\Delta m^2_{ji}L}{4E}\right)  \nonumber \\
&+& 
        2 \sum^3_{j>i} Im\left[
        N^*_{\mu j}N_{ej}N_{\mu i}N^*_{ei}\right] 
        \sin\left(\frac{\Delta m^2_{ji}L}{2E}\right)  .
\end{eqnarray}
And now, instead of the usual unitarity condition for the $3\times 3$
case, we must use the condition given in
Eqs.~(\ref{eq:unita}) and (\ref{eq:NN}), arriving to the expression
\begin{eqnarray}
P_{\mu e} = \alpha_{11}^2 |\alpha_{21}|^2  &-& 4  
        \sum^3_{j>i} Re\left[ 
        N^*_{\mu j}N_{ej}N_{\mu i}N^*_{ei} \right]
        \sin^2\left(\frac{\Delta m^2_{ji}L}{4E}\right)  \nonumber \\
&+& 
        2 \sum^3_{j>i} Im\left[
        N^*_{\mu j}N_{ej}N_{\mu i}N^*_{ei}\right] 
        \sin\left(\frac{\Delta m^2_{ji}L}{2E}\right)  
        . 
\end{eqnarray}
Using Eq.~(\ref{eq:Ndescopm_C1}) one can substitute the values of
$N_{\alpha i}$ in terms of $U_{\alpha i}$ and $\alpha_{ij}$ to obtain
\begin{eqnarray}
P_{\mu e} &=& 
 \alpha_{11}^2|\alpha_{21} |^2 \left(
 1 - 4  
        \sum^3_{j>i} 
        |U_{ej}|^2|U_{ei}|^2
        \sin^2\left(\frac{\Delta m^2_{ji}L}{4E}\right) 
\right) \nonumber \\
 &-& 
 (\alpha_{11}\alpha_{22})^2 4
        \sum^3_{j>i} Re\left[
        U^*_{\mu j}U_{ej}U_{\mu i}U^*_{ei} \right]
        \sin^2\left(\frac{\Delta m^2_{ji}L}{4E}\right)
\nonumber \\
 &+& 
 (\alpha_{11}\alpha_{22})^2 2
        \sum^3_{j>i} Im\left[
        U^*_{\mu j}U_{ej}U_{\mu i}U^*_{ei} \right]
        \sin\left(\frac{\Delta m^2_{ji}L}{2E}\right)
\nonumber \\
 &-& 
4 \alpha_{11}^2 \alpha_{22}
        \sum^3_{j>i} Re \left[ 
        \alpha_{21}|U_{ei}|^2U^*_{\mu j}U_{ej} +\alpha^*_{21}|U_{ej}|^2 U_{\mu i}U^*_{ei} \right]
        \sin^2\left(\frac{\Delta m^2_{ji}L}{4E}\right) 
\nonumber \\
 &+& 
2 \alpha_{11}^2 \alpha_{22}
        \sum^3_{j>i} Im \left[ 
        \alpha_{21}|U_{ei}|^2U^*_{\mu j}U_{ej} +\alpha^*_{21}|U_{ej}|^2 U_{\mu i}U^*_{ei} \right]
        \sin\left(\frac{\Delta m^2_{ji}L}{2E}\right) \, .
\end{eqnarray}
Substituting the terms $U_{\alpha i}$ in our parametrization, and 
neglecting cubic products of $\alpha_{21}$, $\sin\theta_{13}$, and 
$\Delta m^2_{21}$, one obtains

\begin{equation}
P_{\mu e} = 
 (\alpha_{11}\alpha_{22})^2 P^{3\times3}_{\mu e}
+  \alpha_{11}^2 \alpha_{22}|\alpha_{21}|  P^{I}_{\mu e} 
+ \alpha_{11}^2|\alpha_{21} |^2 , 
\label{eq:Pmue}
\end{equation}
where we have denoted the standard three-neutrino conversion
probability $P^{3\times 3}_{\mu e}$ as~\cite{Freund:2001pn,Akhmedov:2004ny,Nunokawa:2007qh} 
\begin{eqnarray}
 P^{3\times 3}_{\mu e} &=& 
 4 \bigg[\cos^2\theta_{12} \cos^2\theta_{23} 
   \sin^2\theta_{12} \sin^2\left(\frac{\Delta m^2_{21}L}{4E_\nu}\right)  \nonumber \\ 
 &+&  \cos^2\theta_{13}\sin^2\theta_{13}
   \sin^2\theta_{23}\sin^2\left(\frac{\Delta m^2_{31}L}{4E_\nu}\right) \bigg]\\
&+& 
   \sin(2\theta_{12})
   \sin\theta_{13}\sin(2\theta_{23})
   \sin\left(\frac{\Delta m^2_{21}L}{2E_\nu}\right)
   \sin\left(\frac{\Delta m^2_{31}L}{4E_\nu}\right) 
   \cos\left(\frac{\Delta m^2_{31}L}{4E_\nu} - I_{123}\right)  , \nonumber 
   \label{eq:Pmue3x3}
\end{eqnarray}
while $P^{I}_{\mu e}$ refers to a term that depends on the $3\times 3$ mixing 
angles, plus an extra CP phase: 
\begin{eqnarray}
P^{I}_{\mu e} & = &
-2 
   \bigg[
   \sin(2\theta_{13}) \sin\theta_{23} 
   \sin\left( \frac{\Delta m^2_{31}L} {4E_\nu}\right)
   \sin\left(\frac{\Delta m^2_{31}L}{4E_\nu} + I_{NP}- I_{123}\right) \bigg]
\nonumber \\ 
  & - &  \cos\theta_{13} \cos\theta_{23} 
  \sin(2\theta_{12}) 
   \sin\left(\frac{\Delta m^2_{21}L}{2E_\nu}\right)
  \sin(I_{NP})
   ,
\label{eq:PmueI}
\end{eqnarray}
with $\text{I}_{123}=-\delta_{CP}= \phi_{12}-\phi_{13}+\phi_{23}$ and
$\text{I}_{\text{NP}}=\phi_{12}-\text{Arg}(\alpha_{21})$.

\begin{figure}
\begin{center}
\includegraphics[width=0.95\textwidth,angle=0]{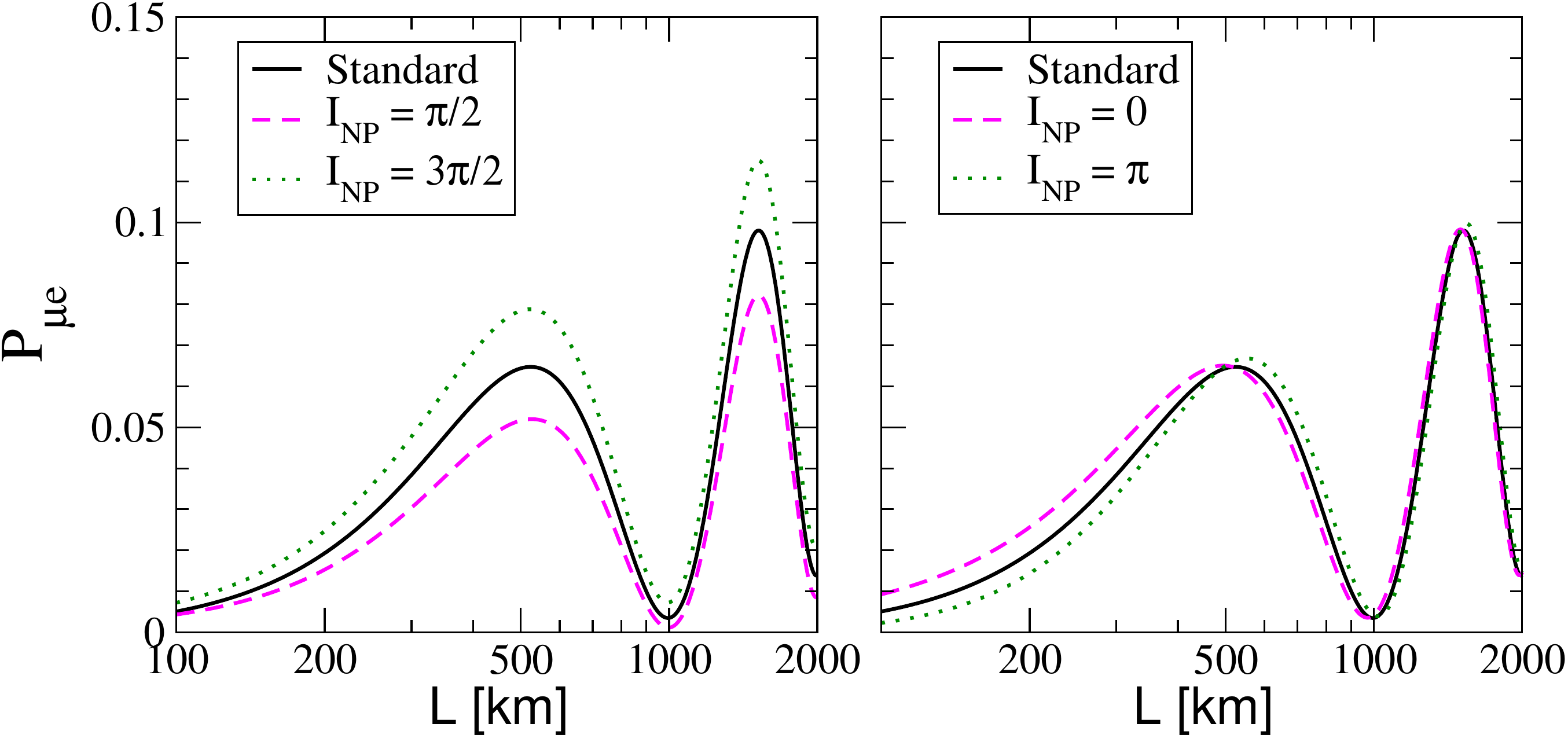}
\caption{Conversion probability for a fixed neutrino energy 
  $E_{\nu}=1$~GeV.  The solid (black) curve shows the standard
  conversion probability, with $\delta = -I_{123} = 3\pi / 2$. The
  non-unitary case is illustrated for $\alpha_{11} = 1$, $\alpha_{22}=
  0.9997$, and $|\alpha_{21}|=0.0264$. In the left panel, two values
  for the new CP phase parameter $I_{NP}$ are considered: $\pi/2$
  (dashed/magenta line) and $3\pi/2$ (dotted/green line), while in the
  right panel we take $I_{NP}=0$ (dashed/magenta line) and $\pi$
  (dotted/green line).}
\label{fig:pmue}
\end{center}
\end{figure}

Notice that the conversion probability depends on just two phases, the
standard one, $I_{123}=-\delta$ and another phase describing the new
physics, $I_{NP}$. This new phase contains the information of the
imaginary part of $\alpha_{21}$, that is, the overall effect of all
the additional phases associated with the heavy states.
Notice that, besides the standard CP term in Eq.~(\ref{eq:Pmue3x3}),
two new CP phase-dependent terms appear; the first involves the
difference between standard and non standard phase: $I_{123}-I_{NP}$,
while the second one depends only on $I_{NP}$. One sees in
Eq.~(\ref{eq:PmueI}) that the first term is proportional to
$\sin\theta_{13}$, while the second one depends on the solar mass
difference $\Delta m^2_{21}$ and, therefore, both terms should be
small. In order to illustrate their impact upon current neutrino data
analysis, we show in Fig.~\ref{fig:pmue} how this new phase parameter
influences the conversion probability. In this figure we compare the
standard three neutrino probability (with a ``best-fit'' phase $\delta
= -I_{123} = 3\pi / 2$), with the case of an additional neutral heavy
lepton with overall contribution given by $\alpha_{11} = 1$,
$\alpha_{22}= 0.9997$, $|\alpha_{21}|=0.0264$, and for the particular
new physics phase parameter of either $\pi/2$ or $3\pi/2$ (left panel)
or $0, \pi$ (right panel). One sees that the effect of the additional
phase in future oscillation appearance experiments could be sizeable
and, depending on the specific value of this new phase, the survival
probability could either increase or decrease.

For the sake of completeness, we also give the expression for the
survival probability $P_{\mu\mu}$:
\begin{equation}
P_{\mu\mu} = \sum^3_i |N_{\mu i}|^2|N_{\mu i }|^2 + 
        \sum^3_{j>i} 2|N_{\mu j}|^2|N_{\mu i}|^2 \cos\left(\frac{\Delta m^2_{ji}}{2E}L\right),
\end{equation}

\begin{equation}
P_{\mu\mu} =  (|\alpha_{21} |^2 + \alpha_{22}^2)^2 - 
4\sum^3_{j>i} |N_{\mu j}|^2|N_{\mu i}|^2 \sin^2\left(\frac{\Delta m^2_{ji}}{4E}L\right),
\end{equation}
\begin{equation}
P_{\mu\mu} = (|\alpha_{21} |^2 + \alpha_{22}^2)^2 - 
        4 \sum^3_{j>i} 
       |\alpha_{21}U_{ej} + \alpha_{22}U_{\mu j}|^2
       |\alpha_{21}U_{ei} + \alpha_{22}U_{\mu i}|^2
       \sin^2\left(\frac{\Delta m^2_{ji}}{4E}L\right),
\end{equation}
so that, neglecting cubic products of $\alpha_{21}$, $\sin\theta_{13}$,
and $\Delta m^2_{21}$, we will obtain
\begin{eqnarray}
P_{\mu\mu} = \alpha_{22}^4 P^{3\times3}_{\mu\mu} 
          + \alpha_{22}^3|\alpha_{21}|   P^{I_1}_{\mu\mu} 
          + 2|\alpha_{21}|^2\alpha_{22}^2 P^{I_2}_{\mu\mu} 
\label{eq:Pmumu3}
\end{eqnarray}
with $P^{3\times3}_{\mu\mu} $, the standard oscillation formula, given by:

\begin{equation}
 \begin{split}
& P_{\mu \mu}^{3\times 3}  \approx 1-4 \left[\cos^2\theta_{23} 
\sin^2\theta_{23}-\cos(2 \theta_{23}) \sin^2\theta_{23} \sin^2{\theta_{13}} 
\right] \,\sin^2 {\left(\frac{\Delta m^2_{31}L}{4E} \right)}\\
&+2 \left[\cos^2\theta_{12} \cos^2\theta_{23} 
\sin^2\theta_{23}- \cos(\text{I}_{123}) \cos \theta_{23} \sin 
(2 \theta_{12}) \sin^3 \theta_{23} \sin{\theta_{13}} 
\right]\,   \sin{\left(\frac{\Delta m^2_{31}L}{2E} \right)} \,\sin{\left(\frac{\Delta m^2_{21}L}{2E} \right)}\\
&-4 \left[\cos^2\theta_{12} \cos^2\theta_{23} 
\sin^2\theta_{23} \,\cos{\left(\frac{\Delta m^2_{31}L}{2E} \right)} +  
\cos^2\theta_{12} \cos^4\theta_{23} 
\sin^2\theta_{12} \right]\,\sin^2{\left(\frac{\Delta m^2_{21}L}{4E}
\right)} \, ,
 \end{split}
\end{equation}
while the extra terms in the oscillation probability are given by:

\begin{equation}
 \begin{split}
 P_{\mu \mu}^{I_1}\approx &-8 \left[
 \sin \theta_{13} \sin{\theta_{23}}\cos(2 \theta_{23})
\cos(\text{I}_{123}-\text{I}_{\text{NP}})  \right] \,\sin^2 
{\left(\frac{\Delta m^2_{31}L}{4E} \right)}\\
&+2\left[
\cos \theta_{23} \sin (2 \theta_{12}) \sin^2\theta_{23} \cos(\text{I}_{\text{NP}})\right]
\,\sin{\left(\frac{\Delta m^2_{31}L}{2E} \right)}
\,\sin{\left(\frac{\Delta m^2_{21}L}{2E} \right)} \, ,
 \end{split}
\end{equation}

\begin{equation}
P_{\mu \mu}^{I_2} \approx 1-2\sin^2\theta_{23}\,\sin^2 
{\left(\frac{\Delta m^2_{31}L}{4E} \right)} .
\end{equation}

\begin{figure}
\begin{center}
\includegraphics[width=0.95\textwidth,angle=0]{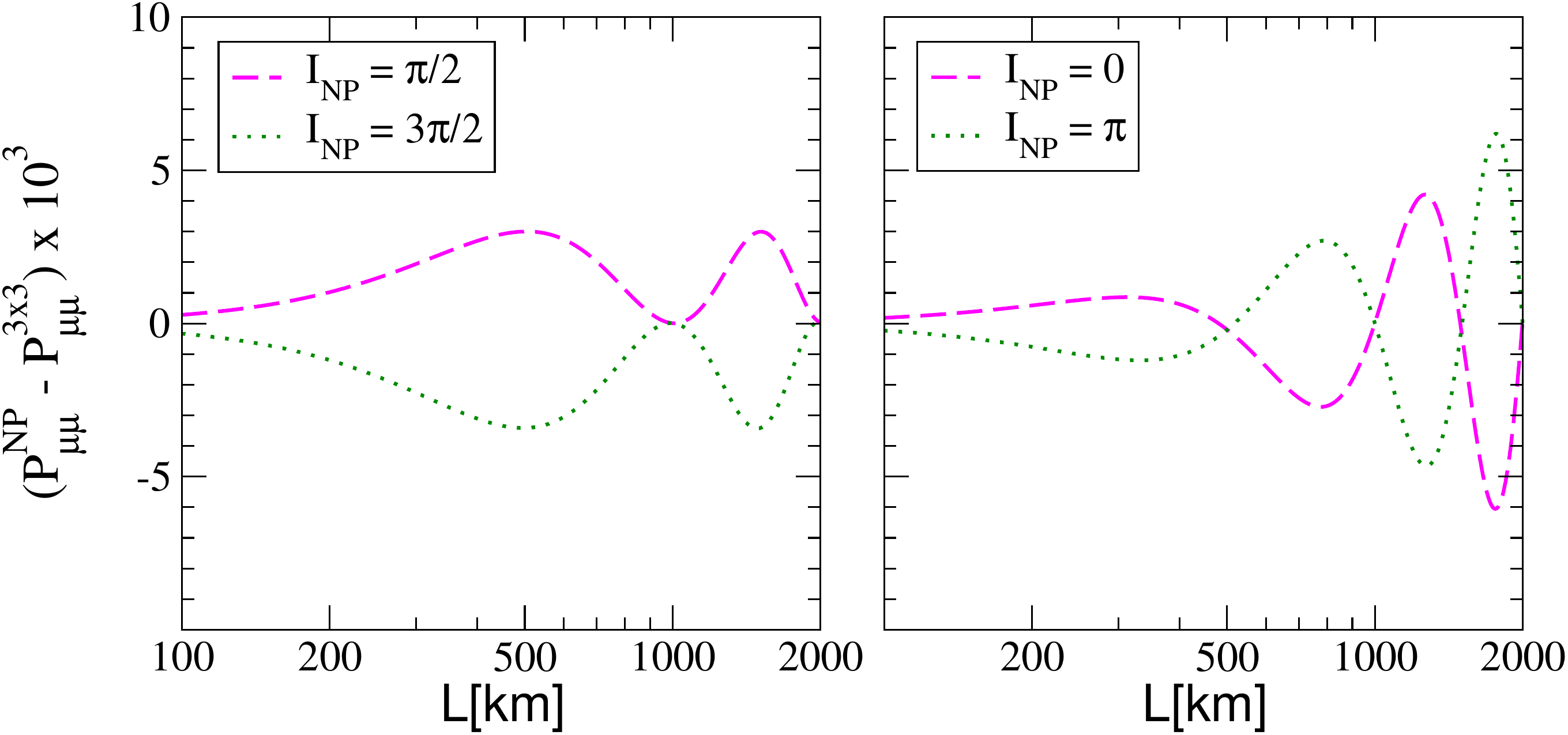}
\caption{Correction to the standard muon neutrino survival probability
  for different values of the new CP phase parameter $I_{NP}$, with
  the remaining parameters fixed as in Fig.~\ref{fig:pmue}.}
\label{fig:pmumu}
\end{center}
\end{figure}

As for the conversion probability, $P(\nu_\mu \to \nu_e$), we also
compute the muon neutrino survival probability and show its behaviour
in Fig.~\ref{fig:pmumu}. As one can see, this disappearance channel is
also sensitive to the new CP phase. The computations were performed
for the same parameter values used in the previous figure, that is,
$\alpha_{11} = 1$, $\alpha_{22}= 0.9997$, $|\alpha_{21}|=0.0264$, and
an overall phase of either $\pi/2$ or $3\pi/2$ as well as $0$ or
$\pi$. The Standard Model phase was fixed to
be $\delta = -I_{123} = 3\pi / 2$).\\


We now turn our attention to oscillations of electron neutrinos or
anti-neutrinos relevant, say, for the description of solar neutrino
experiments, as well as terrestrial experiments using reactors or
radioactive sources.
The electron (anti) neutrino survival probability (in vacuum) is given
by the following expression:
\begin{equation}
P_{ee} = \sum^3_i |N_{ei}|^2|N_{ei}|^2 + 
        \sum^3_{j>i} 2|N_{ej}|^2|N_{ei}|^2 \cos\left(\frac{\Delta m^2_{ji}}{2E}L\right),
\end{equation}
and, using Eq.~(\ref{eq:Ndescopm_C1}), it is easy to see that
$N_{ei}=\alpha_{11}U_{ei}$ which leads to the expression
\begin{equation}
P_{ee} = \alpha^4_{11}\left[ \sum^3_i |U_{ei}|^2|U_{ei}|^2 + 
        \sum^3_{j>i} 2|U_{ej}|^2|U_{ei}|^2 \cos\left(\frac{\Delta m^2_{ji}}{2E}L\right)\right].
\end{equation}
This transforms, in a straightforward way, to the equation 
\begin{equation}
P_{ee} = \alpha^4_{11}\left[1 - \cos^4\theta_{13}\sin^2(2\theta_{12})\sin^2(\Delta_{12})
          - \sin^2(2\theta_{13})\sin^2(\Delta_{13})\right] ,
\label{eq:Pee}
\end{equation} 
with $\Delta_{ij}=\frac{\Delta m^2_{ij}}{4E}L$.  Notice that in this
case, the effect of a neutral heavy lepton will be an overall factor
that accounts for the violation of unitarity: $\alpha_{11}^4$,
unlikely to produce visible effects in oscillations of, say, reactor
neutrinos, given the strong universality restrictions derived in
Fig.~\ref{fig:universality}.
 
For completeness we mention that, should the extra neutrino states be
light enough to take part in oscillations, they could potentially play
a role~\cite{Giunti:2012tn,Giunti:2013aea} in the anomalies reported
by the MiniBooNE collaboration~\cite{Aguilar-Arevalo:2013pmq} or the
reactor neutrino experiments~\cite{Mention:2011rk}. We will not
consider this possibility here.

\section{Bounds from neutrino oscillation experiments}
\label{sec:zero-distance}

From the previous formulas for the oscillation probabilities one sees
that, even at zero distance, the survival and conversion probabilities
differ from one and zero, respectively.  This is a well-known
behaviour and it is a consequence of the effective non-unitarity of
the $3\times 3$ leptonic mixing matrix~\cite{Valle:1987gv}. We can
express these probabilities, for the zero distance case, as
\begin{eqnarray}
P_{ee} &=& \alpha_{11}^4 = [(NN^\dagger)_{11} ]^2
= [1 - (SS^\dagger)_{11}]^2 \, , \nonumber \\
P_{\mu\mu} &=& (|\alpha_{21}|^2+\alpha_{22}^2)^2 = [(NN^\dagger)_{22} ]^2
= [1-(SS^\dagger)_{22}]^2 \, ,  \\
 P_{\mu e} &=& \alpha_{11}^2 |\alpha_{21}|^2  = [(NN^\dagger)_{21} ]^2
= [(SS^\dagger)_{21}]^2 . \nonumber 
\end{eqnarray}

In order to make a quick estimate of the constraints on the new
parameters, we write these expressions in a different way, in order to
compare them with the corresponding expressions for a light sterile
neutrino in the limit of $\Delta m^2_{ij} L / (4E) \gg 1$
($\vev{\sin^2(\Delta m^2_{ij} L/(4E))} = 1/2 $). The result for our
case can be expressed in an analogous way as in the case of extra
light neutrinos~\cite{Kuo:1989qe}:
\begin{eqnarray}
P_{ee} &=& 1 - \frac12 \left[\sin^2\left(2\theta_{ee}\right)\right]_\text{eff} , \nonumber \\
P_{\mu\mu} &=& 1 - \frac12 \left[\sin^2\left(2\theta_{\mu\mu}\right)\right]_\text{eff} , \\
P_{\mu e} &=&  \frac12 \left[\sin^2\left(2\theta_{\mu e}\right)\right]_\text{eff} , \nonumber 
\end{eqnarray}
with 
\begin{eqnarray}
\left[\sin^2\left(2\theta_{ee}\right)\right]_\text{eff}  &=& 2(1-\alpha_{11}^4) , \nonumber \\
\left[\sin^2\left(2\theta_{\mu\mu}\right)\right]_\text{eff} &=& 2[1-(|\alpha_{21}|^2+\alpha_{22}^2)^2 ], \\
\left[\sin^2\left(2\theta_{\mu e}\right)\right]_\text{eff} &=& 2\alpha_{11}^2|\alpha_{21}|^2 . \nonumber 
\end{eqnarray}
We can compare these expressions with the current constraints on light
sterile neutrinos in order to get the following $3\sigma$ limits
~\cite{Giunti:2013aea}
\begin{eqnarray}
\left[\sin^2\left(2\theta_{ee}\right)\right]_\text{eff}   &\leq& 0.2
                                                                 \, , \nonumber \\
\left[\sin^2\left(2\theta_{\mu\mu}\right)\right]_\text{eff} &\leq&
                                                                   0.06
                                                                   \, , \\
\left[\sin^2\left(2\theta_{\mu e}\right)\right]_\text{eff}  &\leq& 1\times 10^{-3} \nonumber \, .
\end{eqnarray}


However, we prefer to use the bound from the NOMAD
experiment~\cite{Astier:2003gs}, since it is the most reliable
constraint on the zero-distance effect (neutrino non-orthonormality
due to heavy neutrino admixture) from neutrino
oscillations. Translated into the parametrization under discussion,
this constraint takes the form

\begin{equation}
\alpha_{11}^2|\alpha_{21}|^2 \leq 0.0007  \, \, \, (90\%\, \text{C.L.})
\end{equation}

If we combine this limit with those coming from universality at
Eqs.~(\ref{eq11}) and (\ref{eq10}), the following 90\% C.L. bounds (1
d.o.f.) are obtained
\begin{eqnarray}
  \centering
\begin{tabular}{ccccc} 
$\alpha_{11}^2\geq 0.989$, & $\,$ & $\alpha_{22}^2\geq 0.999$, & $\,$ & 
                           $|\alpha_{21}|^2 \leq 0.0007$.
\end{tabular}
\end{eqnarray}

\section{Compiling current NHL constraints}

Non-standard features such as unitarity violation in neutrino mixing
could signal new physics responsible for neutrino mass. For example,
they could shed light upon the properties of neutral heavy leptons
such as right-handed neutrinos, which are the messengers of neutrino
mass generation postulated in seesaw schemes. In many such schemes the
smallness of neutrino masses severely restricts the magnitudes of the
expected NHL signatures. However these limitations can be circumvented
within a broad class of low-scale seesaw
realizations~\cite{Mohapatra:1986bd,GonzalezGarcia:1988rw,Akhmedov:1995ip,Akhmedov:1995vm,Malinsky:2005bi,Boucenna:2014zba,Dev:2009aw,Dev:2013oxa,Drewes:2015iva}. For
this reason in this section we will present a compilation of
model-independent NHL limits, which do not require them to play the
role of neutrino mass messenger in any particular seesaw
scheme. Results of this section are not original, but they are
included for logical completeness.

Isosinglet neutrinos have been searched for in a variety of
experiments.  For example, if they are very light they may be emitted
in weak decays of pions and kaons. Heavier ones, but lighter than the
$Z$ boson, would have been copiously produced in the first phase of
the LEP experiment should the coupling be
appreciable~\cite{Dittmar:1989yg,Akrawy:1990zq}. Searches have been
negative, including those performed at the higher, second phase
energies~\cite{Abreu:1996pa}.
\begin{figure}[!h]
\centerline{
\includegraphics[scale=0.5]{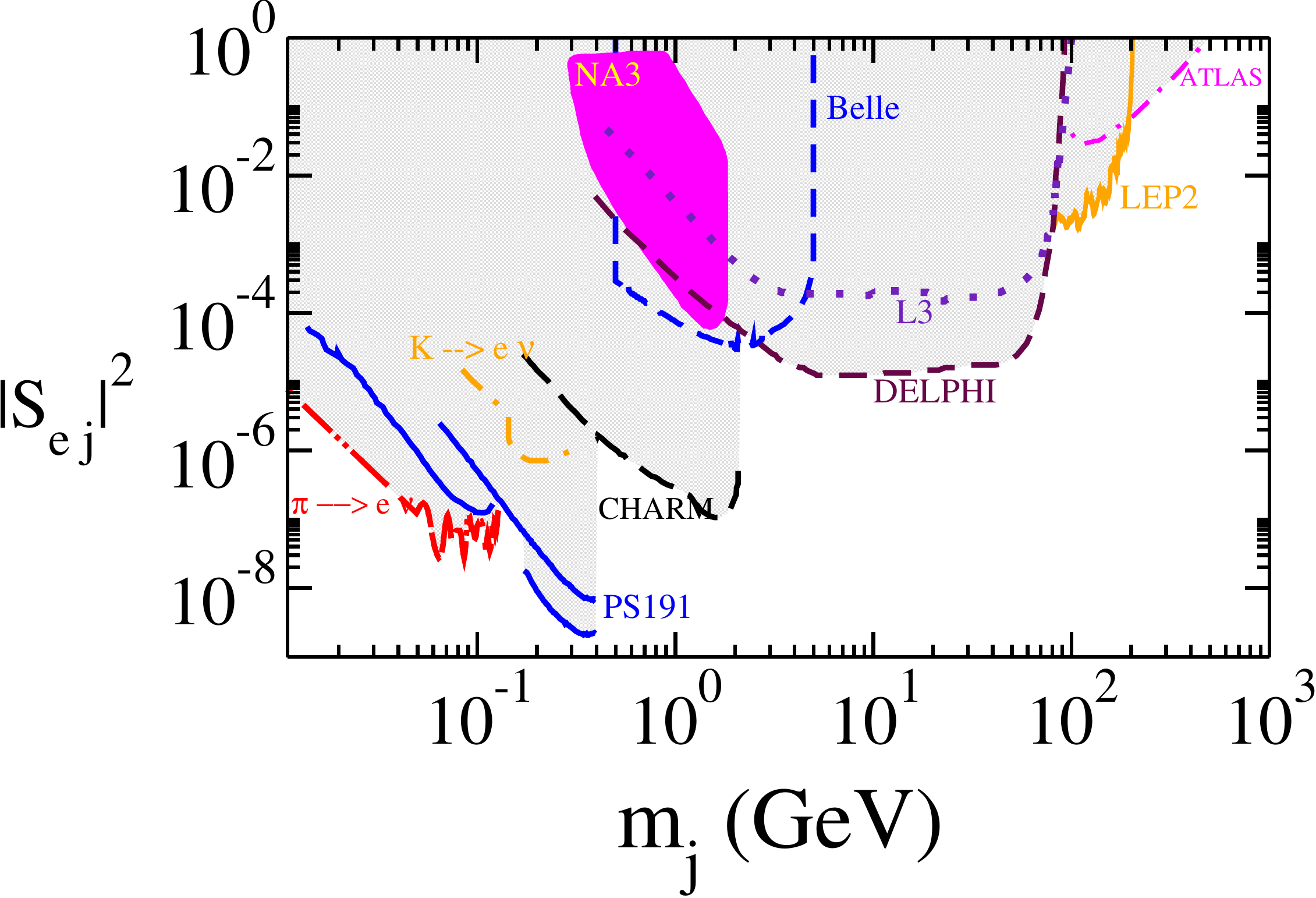}
}
\caption{\label{fig:Kej} Bounds on the component of a heavy isosinglet
  lepton of mass $m_j$ in the electron neutrino.  }
\end{figure}

\begin{figure}
\centerline{
\includegraphics[scale=0.5]{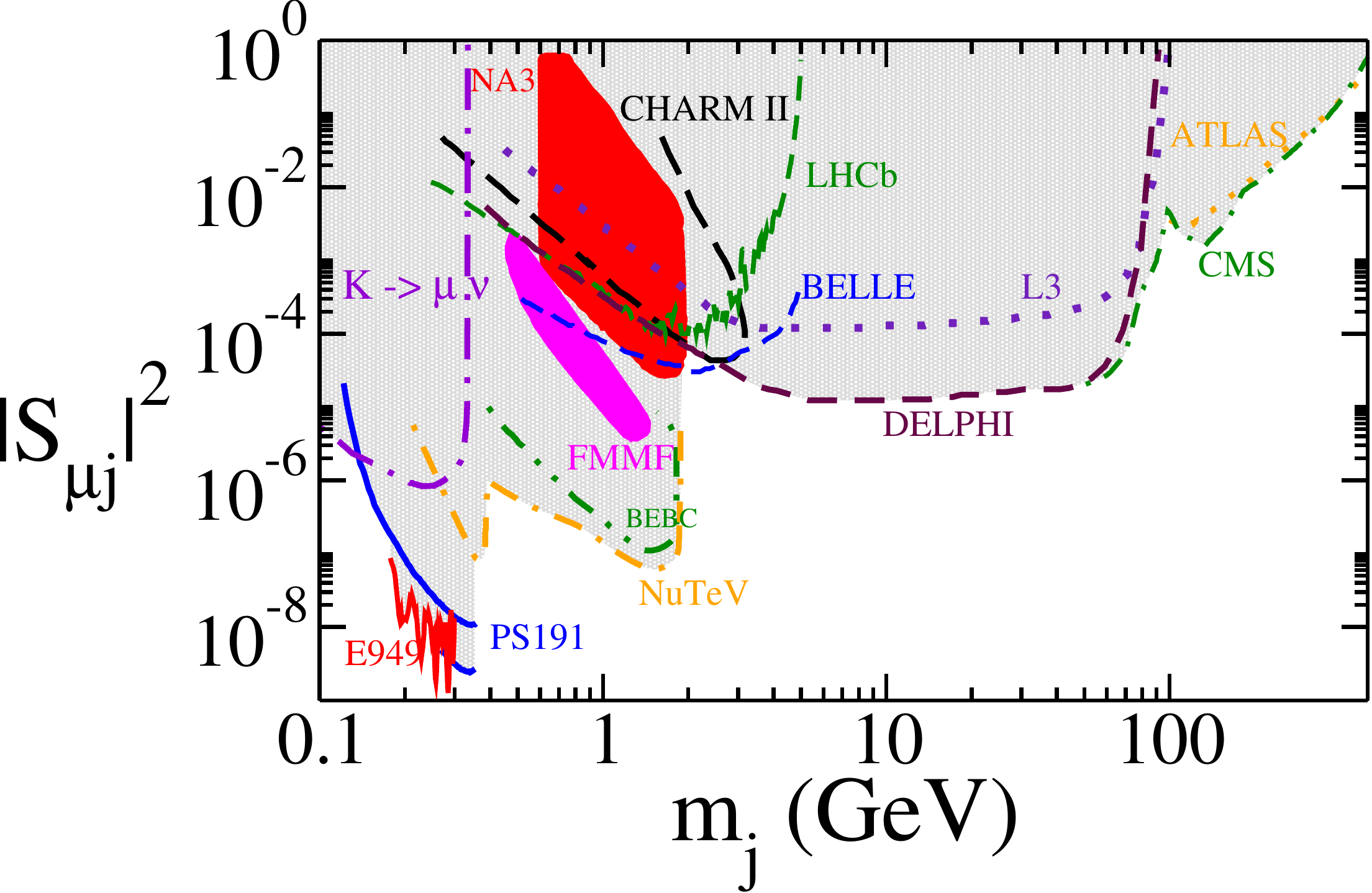}
}
\caption{\label{fig:Kmuj}
Bounds on the component of a heavy isosinglet
  lepton of mass $m_j$ in the muon neutrino.  }
\end{figure}

\begin{figure}[!h]
\centerline{
\includegraphics[scale=0.5]{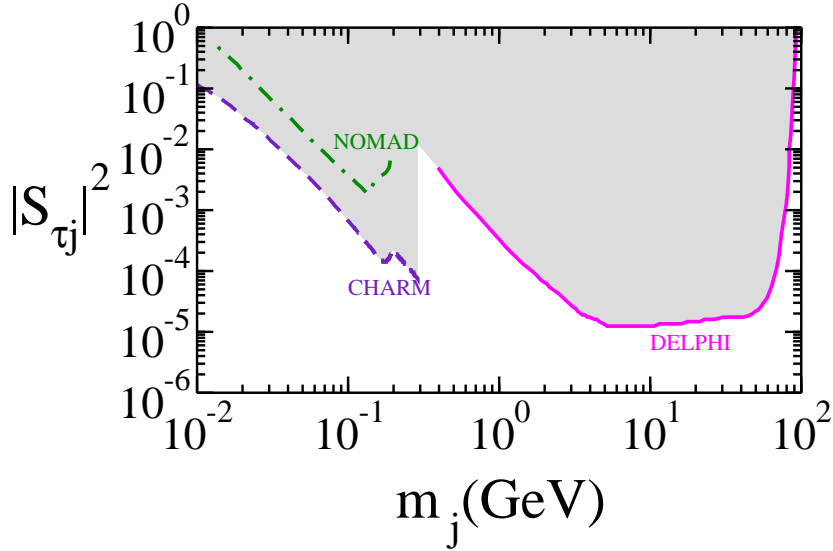}
}
\vglue -2cm
\caption{\label{fig:Ktauj}Bounds on the component of a heavy isosinglet
  lepton of mass $m_j$ in the tau neutrino.  }
\end{figure}
A summary of constraints for the direct production of neutral heavy
leptons is shown in Figs.~\ref{fig:Kej}, \ref{fig:Kmuj} and 
\ref{fig:Ktauj}. In
most cases, experiments have looked for a resonance in a given energy
window, for a given mixing of the additional state, described in this
case by the submatrix $S$ of Eq.~(\ref{eq:ULindner_C1}). Although the
constraints for the mixing in these cases are stronger, in most of the
cases they rely upon extra assumptions on how the heavy neutrino
should decay.

In particular, in Fig.~\ref{fig:Kej}, we summarize the constraints
on $|S_{ej}|^2$ for a mass range from $10^{-2}$ to $10^2$~GeV coming
from the experiments TRIUMF~\cite{Britton:1992pg,Britton:1992xv}
(denoted as $\pi\to e\nu$ and $K\to e\nu$ in the plot),
PS191~\cite{Bernardi:1987ek}, NA3~\cite{Badier:1986xz},
CHARM~\cite{Bergsma:1985is}, Belle~\cite{Liventsev:2013zz}, the LEP
experiments DELPHI~\cite{Abreu:1996pa}, L3~\cite{Adriani:1992pq}, 
LEP2~\cite{Achard:2001qv}, and the recent LHC results from 
ATLAS~\cite{ATLAS:2012yoa,klinger}. 
Future experimental proposals, such as DUNE~\cite{Adams:2013qkq} and
ILC, expect to improve these constraints~\cite{Deppisch:2015qwa}

In Fig.~\ref{fig:Kmuj} we show the corresponding constraints for the
case of the mixing of a neutral heavy lepton with a muon neutrino. In
this case we show the experimental results coming again from PS191,
NA3, and Belle, from the LEP experiments L3, DELPHI, and from the LHC
experiment ATLAS; we also show the bounds coming from
KEK~\cite{Hayano:1982wu,Kusenko:2004qc} (denoted as $K\to \mu\nu$ in
the plot), CHARM II~\cite{Vilain:1994vg}, FMMF~\cite{Gallas:1994xp},
BEBC~\cite{CooperSarkar:1985nh}, NuTeV~\cite{Vaitaitis:1999wq},
E949~\cite{Artamonov:2014urb}, and from the LHC experiments
CMS~\cite{Khachatryan:2015gha} and LHCb~\cite{Aaij:2014aba}.
Finally, for the less studied case of
the mixing of a neutral heavy lepton with a tau neutrino, the known
constraints, coming from NOMAD~\cite{Astier:2001ck},
CHARM~\cite{Orloff:2002de}, and DELPHI~\cite{Abreu:1996pa} are shown
in Fig.~(\ref{fig:Ktauj}).

Heavier neutrinos in the TeV range, natural in the context of
low-scale seesaw, can also be searched for at the LHC. However, within
the standard \SM model such heavy, mainly isosinglet, neutrinos would
be produced only through small mixing effects. Indeed, it can be seen
from Figs.~\ref{fig:Kej},~\ref{fig:Kmuj} and \ref{fig:Ktauj} that
restrictions are rather weak.
In contrast, this limitation can be avoided in extended electroweak
models. In such case a production portal involving extra kinematically
accessible gauge bosons, such as those associated with left-right
symmetric models, can give rise to signatures at high energies, such
as processes with lepton flavour
violation~\cite{AguilarSaavedra:2012fu,Das:2012ii}.

\subsection{Neutrinoless double beta decay}

If neutrinos have Majorana nature, as expected on theoretical grounds,
neutrinoless double beta decay is expected to occur at some
level~\cite{Schechter:1981bd}.  
We start our discussion by reminding the definition of the effective
Majorana neutrino mass~\cite{Deppisch:2012nb},
\begin{equation}
\label{eq:mass-mech}
\vev{m} = |\sum_j (U^{n\times n}_{ej})^2 m_j | ,
\end{equation}
where the sum runs only for the light neutrinos coupling to the
electron and the $W$-boson.  \\

From Eq.~(\ref{eq:Ndescopm_C1}) one sees that, in the presence of the
heavy neutrinos, the three light SM neutrino charged current couplings
will be modified to $U^{n\times n}_{ei} = \alpha_{11}U_{ei}$, with
i=1,2,3, and their contribution to neutrinoless double beta decay will
change correspondingly.
\begin{figure}[!h]
\centerline{
\includegraphics[scale=0.5]{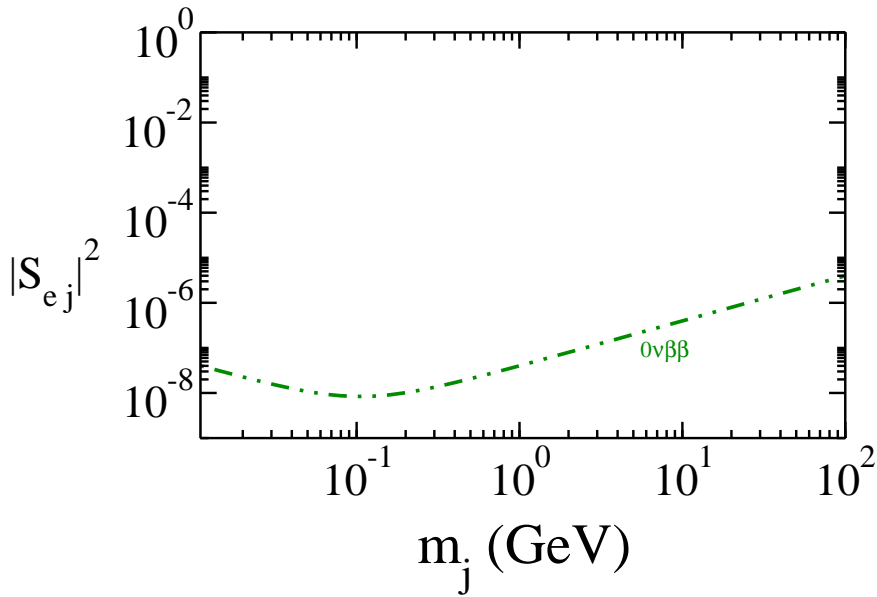}
}
\vglue -2cm
\caption{\label{fig:0nubb} Sensitivity of neutrinoless double beta
  decay to isosinglet mass $m_j$ in the electron neutrino.  }
\end{figure}

Moreover, the heavy states will induce also a short-range or contact
contribution to neutrinoless double beta decay involving the exchange
of the heavy Majorana neutrinos. Since these are \SM singlets they
couple only through the mixing coefficients $S_{ej}$. The general form
of the amplitude is proportional to
\begin{equation}
{\cal A} \propto \frac{m_j}{q^2-m^2_j},
\end{equation}
where $q$ is the virtual neutrino momentum transfer. Clearly there are
two main regimes for this amplitude; for $q^2 \gg m^2_j$, we have
\begin{equation}
{\cal A}_{\rm light} \propto m_j ,
\end{equation}
while for $q^2 \ll m^2_j$ 
\begin{equation}
{\cal A}_{\rm heavy}  \propto \frac{1}{m_j} \, .
\end{equation}
This behaviour can be seen in the corresponding estimated sensitivity
curve shown in Fig. ~(\ref{fig:0nubb}).  This line is obtained for
$^{76}$Ge assuming a single massive isosinglet
neutrino~\cite{Mitra:2011qr}. The change in slope takes place for
masses close to the typical nuclear momentum, around
$100-200$~MeV. Both light and heavy contributions must be folded in
with the appropriate nuclear matrix elements~\cite{Simkovic:2010ka}
whose uncertainties are still large. As a result it is not possible to
probe the indirect NHL effect upon the light neutrino contribution to
the effective mass in Eq.~(\ref{eq:mass-mech}) which amounts to a
multiplicative factor $\alpha_{11}^2$ in the amplitude, a difference
well below current sensitivities.
Notice that, in contrast to bounds discussed in Figs.~\ref{fig:Kej}, 
\ref{fig:Kmuj} and \ref{fig:Ktauj} the restriction from the
neutrinoless double beta decay in Fig.~\ref{fig:0nubb} holds only if
neutrinos have Majorana nature.

\subsection{Charged \lfv}

Virtual exchange of NHLs would also induce charged lepton flavour
violation processes both at low energies~\cite{Forero:2011pc} as well
as in the high energies provided by accelerator
experiments~\cite{Bernabeu:1987gr}. However rates would depend on
additional flavor parameters and upon details on the seesaw mechanism
providing masses to neutrinos.
The possibility of probing it at hadronic colliders such as the LHC
may be realistic in low-scale seesaw models with additional TeV scale
gauge bosons beyond those of the SM gauge structure and with lighter
NHLs~\cite{AguilarSaavedra:2012fu,Das:2012ii,Deppisch:2013cya,Chen:2013foz}.
However we do not consider this possibility any further here because
the corresponding rates depend on very model-dependent assumptions.

\section{Summary}

Simplest seesaw extensions of the Standard Model predict unitarity
deviations in the leptonic mixing matrix describing the charged
current leptonic weak interaction. This is due to the admixture of
heavy isosinglet neutrinos, such as ``right-handed neutrinos'', which
are the ``messengers'' whose exchange generates small neutrino masses.
Low-scale realizations of such schemes suggest that such NHL may be
light enough as to be accessible at high energy colliders such as the
LHC or, indirectly, induce sizeable unitarity deviations in the
``effective'' lepton mixing matrix.
In this paper we used the general symmetric parametrization of lepton
mixing of Ref.~\cite{Schechter:1980gr} in order to derive a simple
description of unitarity deviations in the light neutrino sector. Most
experiments employ neutrinos or anti-neutrinos of the first two
generations. Their description becomes especially simple in our
method, Eq.~(\ref{eq:Ndescopm_C1}), as it involves only a subset of
parameters consisting of three real effective parameters plus a single
CP phase.
We have illustrated the impact of non-unitary lepton mixing on weak
decay processes as well as neutrino oscillations. For logical
completeness we have also re-compiled the current model-independent
constraints on heavy neutrino coupling parameters arising from various
experiments in this notation.
In short, our method will be useful in a joint description of NHL
searches as well as upcoming precision neutrino oscillation studies,
and will hopefully contribute to shed light on the possible seesaw
origin of neutrino mass.

\section*{Acknowledgements}
This work has been supported by the Spanish grants FPA2014-58183-P and
Multidark CSD2009-00064 (MINECO), and PROMETEOII/2014/084 (Generalitat
Valenciana), by EPLANET, and by the CONACyT grant 166639 (Mexico). MT
is also supported by a Ramon y Cajal contract of the Spanish MINECO.
DVF has been supported by the U.S. Department of Energy under award
number DE-SC0003915.

\section{Appendix: neutrino mixing and heavy isosinglets}

As already explained, heavy gauge singlet neutrinos arise naturally in
several extensions of the Standard Model. The general form of the
mixing matrix describing their charged current weak interaction has
been given in~\cite{Schechter:1980gr}. Here we will further develop
the formalism so as to describe not only the couplings of the
additional heavy neutrinos but also their effects in the light
neutrino sector in a convenient but complete way, with no assumptions
about CP conservation.
Using Okubo's notation~\cite{Okubo:1962zzc}, we can construct the
rotation matrix $U^{n\times n}$ as:
\begin{equation}
U^{n\times n}=\omega_{n-1\, n}\:\omega_{n-2\, n}\:\ldots\:\omega_{1\, n}\:\omega_{n-2\, n-1}\:\omega_{n-3\, n-1}\:\ldots\:\omega_{1\, n-1}\:\ldots\:\omega_{2\,3}\:\omega_{1\,3}\:\omega_{1\,2}\label{eq: Ueq} \, ,
\end{equation}
where each $\omega_{ij}\;(i<j)$ stands for the usual complex rotation
matrix in the $ij$ plane~\cite{Rodejohann:2011vc}:
\begin{equation}
\omega_{13}=\left(\begin{array}{ccc}c_{13} & 0 & e^{-i\phi_{13}}s_{13}\\
0 & 1 & 0\\
-e^{i\phi_{13}}s_{13} & 0 & c_{13}
\end{array}\right)\label{eq: om_matrix} ,
\end{equation} 
with ${s}_{ij}=\sin\theta_{ij}$ and  ${c}_{ij}=\cos\theta_{ij}$. 
This matrix can be expressed in general as:
\begin{equation}
(\omega_{ij})_{\alpha\beta}=
   \delta_{\alpha\beta}\sqrt{1-\delta_{\alpha i}\delta_{\beta j}s_{ij}^{2}
                -\delta_{\alpha j}\delta_{\beta i}s_{ij}^{2}}
          +{\eta}_{ij}\delta_{\alpha i}\delta_{\beta j}
          +\bar{\eta}_{ij}\delta_{\alpha j}\delta_{\beta i} \, ,
          \label{eq:om_form}
\end{equation}
where $i<j$ and
$s^2_{ij}=\sin^{2}\theta_{ij},\;{\eta}_{ij}=e^{-i\phi_{ij}}\,\sin\theta_{ij}$
and $\bar{\eta}_{ij}=-e^{i\phi_{ij}}\,\sin\theta_{ij}$, generalizing the
matrix in Eq.~(\ref{eq: om_matrix}) as:
\begin{equation}
\omega_{ij}=\left(\begin{array}{ccccccccc} 1 & 0 &  & \cdots & 0 & \cdots &  &  & 0\\
0 & 1 &  &  &  &  &  &  & \vdots\\
\vdots &  & c_{ij} & \cdots & 0 & \cdots & {\eta}_{ij}\\
 &  & \vdots & \ddots &  &  & \vdots\\
 &  & 0 &  & 1 &  & 0\\
 &  & \vdots &  &  & \ddots & \vdots\\
 &  & \bar{\eta}_{ij} & \cdots & 0 & \cdots & c_{ij} &  & \vdots\\
\vdots &  &  &  &  &  &  & 1 & 0\\
0 &  &  & \cdots & 0 & \cdots &  & 0 & 1
\end{array}\right) \, . \label{eq:om_gen}
\end{equation}
In general, one can decompose Eq.~(\ref{eq: Ueq}) in the following way
\begin{equation}
U^{n\times n}=U^{n-N}\, U^{N}\label{eq:U_n_matrix} ,
\end{equation}
with
\begin{equation}
U^{N}=\omega_{N-1\, N}\:\omega_{N-2\, N}\:\ldots\:\omega_{1\, N}\label{eq:R_N} ,
\end{equation}
\begin{equation}
U^{n-N}=\omega_{n-1\, n}\:\omega_{n-2\, n}\:\ldots\:\omega_{1\, n}\:\omega_{n-1\, n-1}\:\omega_{n-2\, n-1}\:\ldots\:\omega_{1\, N+1}\label{eq:R_n-N} ,
\end{equation}
so that the matrix decomposition will be given by
\begin{equation}
U^{n-N}U^{N}=\left(\begin{array}{cccccccc}\alpha_{11} & 0 & \cdots & 0 & \vdots\\
\alpha_{21} & \alpha_{22} & \ddots & \vdots & \vdots\\
\vdots &  & \ddots & 0 & \vdots &   & S\\
\alpha_{N1} & \cdots &  & \alpha_{NN} & \vdots\\
\cdots & \cdots & \cdots & \cdots & \vdots & \cdots & \cdots & \cdots \\
 &  &  &  & \vdots\\
 &  & V' &  & \vdots &  & T\\
 &  &  &  & \vdots
\end{array}\right)\:\left(\begin{array}{ccccccc}U_{11}^{N} & U_{12}^{N} & \cdots & U_{1N}^{N} & \vdots\\
U_{21}^{N} & U_{22}^{N} &  & \vdots & \vdots\\
\vdots &  & \ddots &  & \vdots & 0 & \\
U_{N1}^{N} & \cdots &  & U_{NN}^{N} & \vdots\\
\cdots & \cdots & \cdots & \cdots & \vdots & \cdots & \cdots  \\
 &  &  &  & \vdots\\
 &  & 0 &  & \vdots & I & \\
 &  &  &  & \vdots
\end{array}\right)\label{eq:RN_decomp},
\end{equation}
which turns out to be very convenient.
The $3\times3$ neutrino mixing matrix, $U^{3\times3}$, determined in
oscillation experiments could be unitary, or it could be just a
non-unitary submatrix of the larger mixing matrix $U^{n\times n}$
described in Eq.~(\ref{eq: Ueq}).  Therefore, when dealing with more
than three neutrinos, we can write $U^{n\times n}$ as the product of two
matrices:
\begin{equation}
U^{n\times n}=U^{NP}\, U^{SM}\label{eq:Udescomp},
\end{equation}
where "$NP$" means "new physics'' and "$SM$" stands for the ``Standard
Model'' matrix, 
\begin{equation}
U^{NP}=\omega_{n-1\, n}\:\omega_{n-2\, n}\:\ldots\:
\omega_{3\, n}\:
\omega_{2\, n}\:
\omega_{1\, n}\:
\omega_{n-2\, n-1}\:\ldots\:\omega_{3\, n-1}\:
\omega_{2\, n-1}\:
\omega_{1\, n-1}\:
\ldots\:\omega_{3\,4}\:\:\omega_{2\,4}\:\omega_{1\,4} \, , \label{eq:R^NP}
\end{equation}
\begin{equation}
U^{SM}=\omega_{2\,3}\:\omega_{1\,3}\:\omega_{1\,2} \, . \label{eq:R^SM}
\end{equation}
The complete $n\times n$ matrix, $U^{n\times n}$, may be written
as~\cite{Hettmansperger:2011bt}
\begin{equation}
U^{n\times n}=\left(\begin{array}{cc} N & S\\
V & T
\end{array}\right)\label{eq:ULindner} ,
\end{equation}
where $N$ is the $3\times3$ matrix with the standard neutrino terms.
From Eq.~(\ref{eq:Udescomp}) one sees that $N$ can always be
parametrized as
\begin{equation}
N=N^{NP}\, U^{3\times3}=\left(\begin{array}{ccc}\alpha_{11} & 0 & 0\\
\alpha_{21} & \alpha_{22} & 0\\
\alpha_{31} & \alpha_{32} & \alpha_{33}
\end{array}\right)\: U^{3\times3}\label{eq:Ndescopm},
\end{equation}
where the zero triangle submatrix characterizes this decomposition.  It
is useful to see how the components $\alpha_{ij}$ of this matrix can
be found. First notice that $\omega_{i\, j}\omega_{k\,l}$ commutes
when $i\neq k,l$ and $j\neq k,l$; therefore, Eq.~(\ref{eq:R^NP}) can
be rewritten as
\begin{eqnarray}
U^{NP}&=&\omega_{n-1\, n}\:\omega_{n-2\, n}\:\ldots\:\omega_{4\, n}\:
\omega_{n-2\, n-1}\:\ldots\:\omega_{4\, n-1}\:\ldots
\omega_{4\, 5}\: \times \nonumber \\
& & \omega_{3\, n}\:
\omega_{2\, n}\:
\omega_{1\, n}\:
\:\omega_{3\, n-1}\:
\omega_{2\, n-1}\:
\omega_{1\, n-1}\:
\ldots\:\omega_{3\,4}\:\:\omega_{2\,4}\:\omega_{1\,4} \, . \label{eq:R^NP_2}
\end{eqnarray}
Clearly, the first line of this equation has no influence in the submatrices 
$N$ and $S$. On the other hand, the second line of the above equation is a
set of products of the form $\omega_{3\, j}
\omega_{2\, j}
\omega_{1\, j}$, each of them having the form: 
\begin{eqnarray}
{\bold\alpha}^j=\omega_{3j}\omega_{2j}\omega_{1j}
&=&\left(\begin{array}{ccccccc} c_{1j} & 0 & 0 & \vdots & 
                       0& {\eta}_{1j} & 0 \\
{\eta}_{2j}\bar{\eta}_{1j} & c_{2j} & 0 & \vdots & 
                       0& {\eta}_{2j} c_{1j} & 0 \\
{\eta}_{3j}c_{2j}\bar{\eta}_{1j} & {\eta}_{3j}\bar{\eta}_{2j} & c_{3j} & \vdots & 
                       0& {\eta}_{3j} c_{2j} c_{1j} & 0 \\
\cdots & \cdots & \cdots & \cdots & \cdots & \cdots & \cdots \\
             0  &  0 & 0 & \vdots & I & 0 & 0 \\ 
c_{3j}c_{2j}\bar{\eta}_{1j} & c_{3j} \bar{\eta}_{2j} & \bar{\eta}_{3j} & \vdots & 0 & 
                        c_{3j} c_{2j} c_{1j}  & 0\\
             0  &  0 & 0 & \vdots & 0 & 0 & I \\ 
\end{array}\right) \nonumber \\
\nonumber \\
&=&\left(\begin{array}{ccccccc} \alpha^j_{11} & 0 & 0 & \vdots & 
                       0& \alpha^j_{1j} & 0 \\
\alpha^j_{21} & \alpha^j_{22} & 0 & \vdots & 
                       0& \alpha^j_{2j} & 0 \\
\alpha^j_{31} & \alpha_{32} & \alpha^j_{33} &\vdots &  
                       0& \alpha_{3j} & 0 \\
\cdots & \cdots & \cdots & \cdots & \cdots & \cdots & \cdots \\
             0  &  0 & 0 & \vdots &  I & 0 & 0 \\ 
\alpha^j_{j1} & \alpha^j_{2j} & \alpha^j_{j3} &\vdots &  0 & 
                        \alpha^j_{jj}  & 0\\
             0  &  0 & 0 &\vdots &  0 & 0 & I \\ 
\end{array}\right) \, .
\label{eq:R^NP_submatrix}
\end{eqnarray}
We can see that the expression for $N^{NP}$ depends only on products of
the type $\alpha^{n}\alpha^{n-1}\cdots\alpha^{5}\alpha^{4}$.  After
performing the multiplication one notes that the diagonal entries of
the matrix $N^{NP}$ are in general given by
\begin{eqnarray}
\alpha_{11} \: = \: \alpha^{n}_{11} \: \alpha^{n-1}_{11} \: \alpha^{n-2}_{11}    
                                \:\cdots \: \alpha^{4}_{11} 
= \: c_{1\, n}\: c_{\,1n-1}c_{1\, n-2}\ldots c_{14}\nonumber \, , \\
\alpha_{22} \: = \: \alpha^{n}_{22} \: \alpha^{n-1}_{22} \: \alpha^{n-2}_{22}    
                                \:\cdots \: \alpha^{4}_{22} 
= \: c_{2\, n}\: c_{\,2n-1}c_{2\, n-2}\ldots c_{24}\nonumber \, ,\\
\alpha_{33} \: = \: \alpha^{n}_{33} \: \alpha^{n-1}_{33} \: \alpha^{n-2}_{33}    
                                \:\cdots \: \alpha^{4}_{33} 
= \: c_{3\, n}\: c_{\,3n-1}c_{3\, n-2}\ldots c_{34}\nonumber \, ,
\end{eqnarray}
while the off-diagonal entries $\alpha_{ij}$ are given as:
\begin{eqnarray}
\alpha_{21} \: &=& \: \alpha^{n}_{21} \: \alpha^{n-1}_{11} 
                                \:\cdots \: \alpha^{4}_{11} + 
\: \alpha^{n}_{22} \: \alpha^{n-1}_{21} 
                                \:\cdots \: \alpha^{4}_{11} + \cdots +
\: \alpha^{n}_{22} \: \alpha^{n-1}_{22} \: \alpha^{n-2}_{22}    
                                \:\cdots \: \alpha^{4}_{21}  \, ,
\nonumber \\
\alpha_{32} \: &=& \: \alpha^{n}_{32} \: \alpha^{n-1}_{22}
                                \:\cdots \: \alpha^{4}_{22} + 
\: \alpha^{n}_{33} \: \alpha^{n-1}_{32} 
                                \:\cdots \: \alpha^{4}_{22} + \cdots + 
\: \alpha^{n}_{33} \: \alpha^{n-1}_{33} 
\: \alpha^{n-2}_{33}    
                                \:\cdots \: \alpha^{4}_{32} \, ,
\nonumber \\
\alpha_{31} \: &=& \: \alpha^{n}_{31} \: \alpha^{n-1}_{11} 
                                \:\cdots \: \alpha^{4}_{11} + 
\: \alpha^{n}_{33} \: \alpha^{n-1}_{31} 
                                \:\cdots \: \alpha^{4}_{11} + \cdots +
\: \alpha^{n}_{33} \: \alpha^{n-1}_{33} 
\: \alpha^{n-2}_{33}    
                                \:\cdots \: \alpha^{4}_{31} 
\nonumber \\
&+&\: \alpha^{n}_{32} ( \: \alpha^{n-1}_{21} \: \alpha^{n-2}_{11}    
                                \:\cdots \: \alpha^{4}_{11} + 
\: \alpha^{n-1}_{22} \alpha^{n-2}_{21} 
                                \:\cdots \: \alpha^{4}_{11} +  \cdots + 
\: \alpha^{n-1}_{22} \alpha^{n-2}_{22} 
                                \:\cdots \: \alpha^{4}_{21} )
\nonumber \\
&+&
\: \alpha^{n}_{33} \: \alpha^{n-1}_{32} 
(\: \alpha^{n-2}_{21} \: \alpha^{n-3}_{11} \: \cdots \alpha^4_{11} + \cdots + 
  \: \alpha^{n-2}_{22} \: \alpha^{n-3}_{22} \: \cdots \alpha^4_{21}) + \cdots 
\nonumber \\
&+&\: \alpha^{n}_{33} \: \alpha^{n-1}_{33} \: \alpha^{n-2}_{32} 
(\: \alpha^{n-3}_{21} \: \alpha^{n-4}_{11} \: \cdots \alpha^4_{11} + \cdots + 
  \: \alpha^{n-3}_{22} \: \alpha^{n-4}_{22} \: \cdots \alpha^4_{21}) + \cdots  
\nonumber \\
&+& \: \alpha^{n}_{33} \: \alpha^{n-1}_{33} \: \alpha^{n-2}_{33}    
                                \:\cdots \: \alpha^{5}_{32} \: \alpha^{4}_{21} \, ,
\end{eqnarray}
or, more explicitly,
\begin{eqnarray}
\alpha_{21} & = &
    c_{2\, n}\: c_{\,2n-1}\ldots c_{2\, 5}\:{\eta}_{24}\bar{\eta}_{14}\: +\: 
    c_{2\, n}\: \ldots c_{2\, 6}\:{\eta}_{25}\bar{\eta}_{15}\:c_{14} +\: 
    \ldots\:+ {\eta}_{2n}\bar{\eta}_{1n}\:c_{1n-1}\:c_{1n-2}\:\ldots\:c_{14} 
\nonumber \, ,\\
\alpha_{32} & = &
    c_{3\, n}\: c_{\,3n-1}\ldots c_{3\, 5}\:{\eta}_{34}\bar{\eta}_{24}\: +\: 
    c_{3\, n}\: \ldots c_{3\, 6}\:{\eta}_{35}\bar{\eta}_{25}\:c_{24} +\: 
    \ldots\:+ {\eta}_{3n}\bar{\eta}_{2n}\:c_{2n-1}\:c_{2n-2}\:\ldots\:c_{24} 
\nonumber \, , \\
\alpha_{31} & = &
    c_{3\, n}\: c_{\,3n-1}\ldots c_{3\, 5}\:{\eta}_{34}c_{2\,4}\bar{\eta}_{14} + 
    c_{3\, n}\ldots c_{3\, 6}\:{\eta}_{35}c_{2\,5}\bar{\eta}_{15}\:c_{14} + 
    \ldots + {\eta}_{3n}c_{2\, n}\:\bar{\eta}_{1n}\:c_{1n-1}\:c_{1n-2}\ldots c_{14} 
\nonumber \\
    & + & 
    c_{3\, n}\: c_{\,3n-1}\ldots c_{3\, 5}\:{\eta}_{35}\bar{\eta}_{25}
    {\eta}_{24}\bar{\eta}_{14}\: +\: 
    c_{3\, n}\: \ldots c_{3\, 6}\:{\eta}_{36}\bar{\eta}_{26}   c_{2\, 5}\:
                       {\eta}_{24}\bar{\eta}_{14} \nonumber \\ & + & \ldots\: + 
    {\eta}_{3n}\bar{\eta}_{2n}{\eta}_{2n-1}\bar{\eta}_{1n-1}c_{1n-2}\ldots c_{14} \, .
\label{eq:alfa_crossed}
\end{eqnarray}

With these formulas, and the known expression for $U^{3\times3}$, we
already have the explicit description of Eq.~(\ref{eq:Ndescopm}) for
any number of extra neutrino states.
Before concluding this appendix, we would like to remark that the
position of the three off-diagonal zeros in $N^{NP}$ was chosen to
conveniently make the matrix lower triangular. This simplifies the
form of the non-unitary lepton mixing matrix describing most
situations of phenomenological interest, involving solar, atmospheric,
reactor and accelerator neutrinos.
By choosing alternative factor-orderings, one can have different
parameterizations, with the zeros located at different off-diagonal
entries.

\subsection*{Application to 3 + 1 seesaw scheme}

We will conclude this appendix by showing the expressions for
$\alpha_{ij}$ in the case of one and three additional neutrinos.
For the case of just one additional neutrino, the mixing matrix is
given by
\begin{equation}
U^{4\times4}=\left(\begin{array}{cc}N_{3\times3} & S_{3\times1}\\
T_{1\times3} & V_{1\times1}
\end{array}\right)\label{eq:U4} .
\end{equation}
The corresponding expressions for the parameters $\alpha_{ij}$ will be 
given by 
\begin{eqnarray}
\alpha_{11} & = & c_{14}\nonumber \, ,\\
\alpha_{22} & = & c_{24}\nonumber \, , \\
\alpha_{33} & = & c_{34}\nonumber \, ,\\
\alpha_{21} & = & {\eta}_{24}\,\bar{\eta}_{14} \label{eq:alpha31} \, ,\\
\alpha_{32} & = &{\eta}_{34}\,\bar{\eta}_{24}\nonumber \, ,\\
\alpha_{31} & = & {\eta}_{34}\, c_{24}\, \bar{\eta}_{14}\nonumber\, .
\end{eqnarray}

\subsection*{Application to 3 + 3 seesaw scheme}

In this case the full mixing matrix will have the following structure
\begin{equation}
U^{6\times6}=\left(\begin{array}{cc}N_{3\times3} & S_{3\times3}\\
T_{3\times3} & V_{3\times3}
\end{array}\right)\label{eq:Umatrix_6} \, .
\end{equation}
with the $\alpha$ parameters given by
\begin{eqnarray}
\alpha_{11} & = & c_{16}\, c_{15}\, c_{14}\nonumber \, ,\\
\alpha_{22} & = & c_{26}\, c_{25}\, c_{24}\nonumber \, ,\\
\alpha_{33} & = & c_{36}\, c_{35}\, c_{34} \nonumber \, ,\\
\alpha_{21} & = & {\eta}_{26}\,\bar{\eta}_{16}\, c_{15}\,c_{14} \; +
\; c_{26}\,{\eta}_{25}\,\bar{\eta}_{15}\, c_{14} \; +
\; c_{26}\,c_{25}\,{\eta}_{24}\, \bar{\eta}_{14}  \label{eq:alfa_3+3} \, , \\
\alpha_{32} & = & c_{36}\,c_{35}\,{\eta}_{34}\,\bar{\eta}_{24} \; +
\; c_{36}\,c_{35}\,\bar{\eta}_{25}\, c_{24} \; +
\; {\eta}_{36}\,\bar{\eta}_{26}\,c_{25}\, c_{24}  \nonumber \, ,\\
\alpha_{31} & = & c_{36}\, c_{35}\, c_{34} \, 
   {\eta}_{34}\,  c_{24} \, \bar{\eta}_{14} \;
+ \; c_{36}\, {\eta}_{35} \,  c_{24} \, \bar{\eta}_{15}\, c_{14} \;
+ \; {\eta}_{36}\,  c_{26} \, \bar{\eta}_{16} \, c_{15}\, c_{14} \nonumber 
\\
& + & c_{36}\, {\eta}_{35}\, \bar{\eta}_{25} \, {\eta}_{24}\, \bar{\eta}_{14} \;
+ \; {\eta}_{36}\, \bar{\eta}_{26}\, c_{25} \, {\eta}_{24}\, \bar{\eta}_{14} \;
+ \; {\eta}_{36}\, \bar{\eta}_{26}\, {\eta}_{25} \, \bar{\eta}_{15}\,c_{14} \nonumber \, .
\end{eqnarray}

\bibliographystyle{unsrt}

\end{document}